\documentclass[12pt]{article}

\usepackage{amssymb,amsmath}
\usepackage{array}

\makeatletter
\def\alt{\mathrel{\mathpalette\gl@align<}}
\def\agt{\mathrel{\mathpalette\gl@align>}}
\def\gl@align#1#2{\lower.6ex\vbox{\baselineskip\z@skip\lineskip\z@
\ialign{$\m@th#1\hfil##\hfil$\crcr#2\crcr\sim\crcr}}}
\makeatother

\begin{document}
\begin{flushright}
%{\tt hep-ph/yymmnnn}\\
MIFP-09-08 \\
OSU-HEP-09-02\\
 February, 2009
\end{flushright}
\vspace*{1.0cm}

\begin{center}
\baselineskip 20pt
{\Large\bf
Three family unification in higher dimensional models
%\scriptsize(tentatilve)
}

\vspace{1cm}

{\large
Yukihiro Mimura$^a$ and S. Nandi$^b$
}
\vspace{.5cm}

{\baselineskip 20pt
\it
$^a$Department of Physics, Texas A\&M University,
College Station, TX 77843-4242, USA \\
\vspace{2mm} $^b$Department of Physics, Oklahoma State University, \\
and Oklahoma Center for High Energy Physics, Stillwater, OK 74078,
USA } \vspace{.5cm}

\vspace{1.5cm}
{\bf Abstract}
\end{center}

In orbifold models, gauge, Higgs and the matter fields can be
unified in one multiplet from the compactification of higher
dimensional supersymmetric gauge theory. We study how three families
of chiral fermions can be unified in the gauge multiplet. The bulk
gauge interaction includes the Yukawa interactions to generate
masses for quarks and leptons after the electroweak symmetry is
broken. The bulk Yukawa interaction has global or gauged flavor
symmetry originating from the $R$ symmetry or bulk gauge symmetry,
and the Yukawa structure is restricted. When the global and gauged
flavor symmetries are broken by orbifold compactification, the
remaining gauge symmetry which contains the standard model gauge
symmetry is restricted. 
The restrictions from the bulk flavor symmetries can provide
explanations of fermion mass hierarchy.

%Due to their restrictions, hierarchy of
%fermion masses can be explained by suitable choices of the number of
%the extra dimensions and their shapes.

\thispagestyle{empty}

%\bigskip
\newpage

\addtocounter{page}{-1}

\section{Introduction}
\baselineskip 18pt

The standard model (SM) is  very well established to describe the
physics below the electroweak scale. Theoretically, it is expected
that there exists a model beyond SM. The conceptual motivation to
consider the model beyond SM is to understand the variety of
particles as well as the parameters in SM. In fact, the content of
particles in SM is a collection of widely disparate fields: gauge
bosons coming in three factors (color, weak and hypercharge), three
replicated families of chiral fermions coming in many different
representations for quarks and leptons ($q$, $u^c$, $d^c$, $\ell$
and $e^c$), and a scalar Higgs boson to break electroweak symmetry
and give masses to the chiral fermions. %, as well as the gauge bosons.
This brings lots of parameters in SM: three gauge couplings, masses
and mixings for the quarks and leptons, and a Higgs mass and a Higgs
coupling.
%Especially, the Higgs boson is not observed yet
%and the Higgs sector is not known too much.
%One of the key motivations to consider the model beyond SM
%is to understand the electroweak symmetry breaking in the Higgs sector.
The Higgs scalar has a quadratic divergence
in its mass squared, and thus the electroweak scale
is not stable quantum mechanically if the cutoff scale is very high
such as the Planck scale.
Therefore, there must exist a theory beyond SM around the TeV scale.
%relating to the Higgs sector.
Besides, the masses and mixings of the quarks and leptons originate
from the Yukawa couplings with the Higgs boson, which are the most
of the parameters in SM. In such a sense, the nature of the Higgs
boson is a key ingredient to go beyond the standard model. It is
expected that the Higgs boson will be found at the Large Hadron
Collider (LHC) experiment.

%Supersymmetry (SUSY) is the most promising candidate
%to construct a model beyond SM.
%The quadratic divergence of the Higgs boson mass squared
%is canceled due to the unification of couplings of
%fermions and bosons.
%Therefore, the electroweak scale is stable once it is generated,
%and the model at ultra high energy scale can be extrapolated
%from the weak scale physics.
%Although the particles get doubled naively in the SUSY extension,
%these new SUSY particles around the weak scale
%add an additional attraction
%in minimally extended SUSY standard model (MSSM).
%The measured three gauge couplings can be unified at a scale
%through renormalization group running.
%That leads us an idea of grand unified theory (GUT) \cite{Georgi:1974sy,Dimopoulos:1981zb}
%that the gauge symmetry of the SM
%is unified in a simple group, such as $SU(5)$ and $SO(10)$.

%The quarks and leptons can be unified in a larger
%representation, especially in $SO(10)$ model.
%As a result, masses and mixings of quarks and leptons
%can be investigated in unified picture,
%though the style of unification is not unique.
%However, Higgs sector may become complicate in GUT.
%Naively, the Higgs doublet in SM is also unified in
%a larger representation.
%Then, there should be colored Higgs particles,
%which may cause a rapid proton decay rate rather
%than the current experimental bounds
%in SUSY GUT \cite{Sakai:1981pk}.
%Therefore, the colored Higgs fields should be heavy,
%or the couplings for quarks and leptons needs to be
%forbidden.

The idea of extra dimensions is an attractive candidate
to build a model beyond SM.
%though there is no experimentally supported signal yet.
Kaluza and Klein (KK) showed that it is possible to
interpret electromagnetism as the effect of gravity
in five dimensions under certain projections.~%\cite{Kaluza:tu}.
After dimensional reduction, a vector/tensor field in some higher
dimensional compactified space decomposes into separate scalar,
vector and tensor components in four dimensions (4D).
The left- and right-handed Weyl fermions in 4D are unified
in higher dimensional fermions.
The idea of extra dimensions has not been treated
in phenomenological issues, but it becomes fashionable
since it may explain the large scale hierarchy \cite{Antoniadis:1990ew}.
The idea of compactification is also applied
to break symmetries by orbifold boundary conditions
\cite{Kawamura:1999nj,Hall:2001pg,Mimura:2002te}.
%When it is applied in GUT models,
%colored Higgs particles can be projected out,
%while Higgs doublet can be light as a zero mode.
%The proton decay operator via the colored Higgs fields
%is forbidden in the model~\cite{Kawamura:1999nj}.
Though the gauge symmetry is explicitly broken by the orbifold
conditions, the gauge couplings can be unified when the brane
localized gauge interaction is suppressed by a large volume of the
extra dimensions \cite{Hall:2001pg}. In such fashions, the idea of
gauge-Higgs unification~\cite{Manton:1979kb} is revived
\cite{Dvali:2001qr}. The scalar Higgs fields can be unified with the
gauge fields in some higher dimensional vector fields. In a simple
orbifold boundary condition, the gauge symmetry is broken since the
broken generators of gauge bosons for 4D coordinates are projected
out. At that time, the broken generators for extra dimensional
components can have massless modes, and the Wilson line operator can
be identified as the Higgs bosons to break the symmetry remained in
4D \cite{Hosotani:1983xw}. Interestingly, this idea is compatible to
extend the SM gauge symmetry to a larger gauge group such as in
grand unified theories (GUT). Besides, the mass of the Higgs scalar
is forbidden by gauge invariance, and thus it can remain light at
low energy when supersymmetry (SUSY) is combined in the model.

The interesting consequence of the gauge-Higgs unification
is that the Yukawa interaction can originate from
the gauge interaction when fermions are also
higher dimensional bulk fields~\cite{Burdman:2002se,Gogoladze:2003bb}.
Actually,
the 4D zero modes of fermions can be chiral in orbifold projections,
and
the higher dimensional extension of the fermion kinetic term
with covariant derivative,
$\bar\psi \gamma^\mu (\partial_\mu - i g A_\mu) \psi$,
includes Yukawa coupling when the gauge fields with higher dimensional
components are identified as Higgs fields.
In the left-right symmetric construction of the model~\cite{Shafi:2002ck},
%from 5D ${\cal N} =1$ SUSY $S^1/\mathbb{Z}_2$ orbifold,
the matter representation to realize the gauge-Yukawa unification
can be much simpler than that of the SM construction,
and the actual unification of gauge and Yukawa coupling constants
can be realized~\cite{Gogoladze:2003bb}.
In the models
of 5D ${\cal N} =1$ SUSY $S^1/\mathbb{Z}_2$ orbifold
with bulk gauge symmetries such as $SO(11)$ and $SU(8)$,
which break down to Pati-Salam (PS) symmetry group
 $G_{\rm PS} = SU(4)_c \times SU(2)_L \times SU(2)_R$ \cite{Pati:1974yy} in 4D,
matter fields are unified in hypermultiplets, and
 all three gauge couplings and third generation
Yukawa couplings (top, bottom, tau and Dirac tau neutrino)
can be unified.
Then, the Yukawa couplings at the weak scale can be calculated
assuming that the threshold corrections are small.
Consequently, we can predict
the top quark mass as well as $\tan\beta$,
which is a ratio of Higgs vacuum expectation values (VEVs) for
up- and down-type Higgs bosons.
Actually, the prediction of the top quark mass can agree
with the experiment if we take into account the threshold corrections
\cite{Gogoladze:2003pp}.
If we say it inversely, we can survey if the unification of gauge
and Yukawa couplings is really realized in the future since the LHC
and ILC experiments will provide us more accurate measurement of the
Yukawa couplings above TeV scale. It is important that the
unification of the gauge and Yukawa coupling constants can be a
signal of extra dimensions at ultra high energy scale. Therefore, we
should investigate models in which gauge and Yukawa unification can
happen.

In SUSY extensions,
the matter fermions can be unified in higher dimensional gauge
multiplets \cite{Watari:2002tf,Li:2003ee},
especially when the model consists of $N=4$ vector multiplet in 4D language
such as in 6D ${\cal N}=(1,1)$ SUSY.
Interestingly, three replications of family can be obtained
in the $T^2/\mathbb{Z}_3$ orbifold \cite{Watari:2002tf}.
The three families originate from the three chiral supermultiplets
in the $N=4$ gauge multiplet.
Since the number of chiral multiplet is maximally three
in 4D, it may explain why the family is three times replicated.

The hypermultiplet which is adjoint representation
under the bulk gauge symmetry in 5D ${\cal N}=1$ SUSY $S^1/\mathbb{Z}_2$ orbifold model
can be incorporated into the gauge multiplet in 6D ${\cal N}=(1,1)$
SUSY orbifold models.
%Two of the authors (I.G. and Y.M.) and S. Nandi
In Ref.\cite{Gogoladze:2003ci},
it is found that all matter species for one family and Higgs doublets
as well as gauge fields in SM can be unified
in 6D ${\cal N} = (1,1)$ SUSY $SU(8)$ gauge multiplets
with $T^2/\mathbb{Z}_6$ orbifold.
The three gauge couplings and the third generation Yukawa couplings can be also unified
in the model.
The other gauge groups \cite{Gogoladze:2003yw,Gogoladze:2003cp}
and other extensions including seven dimensional
models \cite{Gogoladze:2005az,Gogoladze:2006ps}
are also considered.
Since no other bulk matter fields can be introduced,
the model can explain why only third family is heavy.
%
%
%The deficit of this type of models
%is that the selection of the $\mathbb{Z}_6$ discrete charge
%is artificial to identify the matter and Higgs fields
%as zero modes.
%The first and second families are just brane fields
%to cancel brane localized gauge anomalies.
%So, the Yukawa couplings for the first and second families
%are suppressed by a large volume factor,
%but there is no stringent reason that
%the mass of the first family is hierarchically small.
%Apparently, the masses and mixings are just given by hand.
%When gauge symmetry is extended such as $SO(16)$,
%other families can be included in the vector multiplet \cite{Gogoladze:2003yw},
%though the discrete charge assignment is more complicate.
%
%
If we choose $T^2/\mathbb{Z}_3$ orbifold in the $SU(8)$ model,
%on the other hand,
the discrete charge assignment is simple and almost unique
when $N=1$ SUSY remains at 4D.
In the 6D $SU(8)$ $T^2/\mathbb{Z}_3$ orbifold model, thus,
three families and Higgs fields as well as
gauge fields are naturally unified in one multiplet \cite{Gogoladze:2007nb}.

%However, the Yukawa coupling matrix is antisymmetric
%originating from the fact that the chiral superfields
%in the gauge multiplet are
%adjoint representation under the bulk gauge symmetry.
%Then, as a result, after the electroweak symmetry is broken,
%two of the families have degenerate masses
%and one of the families is massless.
%It is good that the one of the family is massless
%since that can explain why the first family is hierarchically light.
%However, the degeneracy for the second and third families
%is not phenomenologically viable at all.
%One may introduce the brane localized interaction to make
%one of the mass eigenvalues small by cancellation.
%The cancellation is not only unnatural, but also
%inconsistent with the volume suppression of the brane interactions.
%If we don't take into account the volume suppression,
%the gauge coupling unification is also disturbed by the brane
%gauge couplings.
%If we project out the unwanted second generation by orbifold
%boundary condition,
%it results in the $T^2/\mathbb{Z}_6$ model.

In this paper,
we will study the cases
where there are just three chiral families as zero modes
of the bulk fields
in the higher dimensional orbifold models.
We investigate the Pati-Salam branch of the $E_8$ and its subgroup
to break the symmetry by orbifold compactification, and to extract
the three chiral families as zero modes.
In fact, the adjoint representation of $E_8$
contains maximally four families of matters
as decomposed representation under the Pati-Salam symmetry.
There are three chiral superfields in the bulk
for the higher dimensional models,
and thus, maximally twelve families are contained in the bulk.
By orbifolding, many of the component representations are
projected out, and we choose the orbifold charge assignment
to extract just three chiral families as zero modes.
We will show that
there are five cases for the charge assignments to obtain the
three chiral families,
and give examples of the discrete charge assignments.
%
%In some cases, we obtain the both left- and right-handed
%families are in the bulk.
%
%We also consider the case where the number of left-handed families is three,
%but the number of right-handed families is one or two in the bulk fields.
%The deficits are made up by the brane-localized fields.
%Even in the latter case, the number of the families can come from the
%zero modes of the bulk fields.
%
The Yukawa interaction is generated from the bulk gauge interaction,
and we will classify the structure of the mass eigenvalues in the three-family models.
In some cases, global or gauged flavor symmetry remains in the Yukawa interaction
due to the bulk gauge interaction.
The flavor symmetries originates from the $R$ symmetry or gauge symmetry in the bulk.
In such cases, two of the eigenvalues are degenerate.
In one case, the flavor symmetry is broken by the orbifold projection, and
the eigenvalues are not degenerate, but there is a restriction in the 4D gauge symmetry
which contains Pati-Salam branch.
We study the trinification branch in the typical case to obtain a hierarchical structure
of the fermion masses.
We also consider the case in which SM gauge symmetry $G_{\rm SM} = SU(3)_c \times SU(2)_L \times U(1)_Y$
 remains in 4D.

This paper is organized as follows: In section 2, we introduce the
orbifold models which we consider in this paper. In section 3, we
investigate the Pati-Salam branch from the $E_8$ group and its
subgroups. The $\mathbb{Z}_n$ charges are assigned to the decomposed
fields under the Pati-Salam symmetry in order break the bulk gauge
symmetry down to the Pati-Salam symmetry. In section 4, we will show
the five cases to obtain the three chiral families as zero modes
from the bulk fields, and give examples of the orbifold charge
assignments. In section 5, we make a comment to develop the example
to build a model. In section 6, we consider the trinification
branch, as well as the SM decomposition. Section 7 is devoted to the
conclusions.

\section{Gauge, Higgs and matter unification}

In this section, we will briefly study the higher dimensional orbifold model,
which is used to construct a model where
gauge, Higgs and three families of matter
are unified in the higher dimensional SUSY gauge multiplet.
%The bulk gauge multiplet transforms in the adjoint of the
%gauge group $G=SU(8)$.
We will consider 10D ${\cal N}=1$ SUSY model
to describe the theory generally, but 8D model can be also considered.
Actually, the model examples which we will see later can be made in 8D orbifold.
We consider the extra dimensions are compactified over a flat
$T^2/\mathbb{Z}_{n_1} \times T^2/\mathbb{Z}_{n_2} \times T^2/\mathbb{Z}_{n_3}$
orbifold.
The formalism of the higher dimensional models can be
seen in Ref.\cite{Marcus:1983wb}.

From a 4D point of view, the 10D ${\cal N}=1$ gauge multiplet
is recognized as one $N=4$ multiplet which consists of one ${N}=1$
vector superfield $V$ and three chiral superfields $\Sigma_i$ $(i=1,2,3)$.
The scalar components of the chiral superfields $\Sigma_1, \Sigma_2$, and $\Sigma_3$
are
$A_5 - i A_6$, $A_7-i A_8$, and $A_9 - i A_{10}$, respectively.
We define the extra dimensional coordinates as $z_1 = x_5+i x_6$, $z_2 = x_7 + i x_8$, and $z_3 = x_9+i x_{10}$.
The orbifold transformations ${\bf R}_i$ are $z_i \to \omega z_i$,
where $\omega = e^{2\pi i/n_i}$.
The transformation ${\bf R}_i$ can also act on the internal symmetry of the Lagrangian.
The internal symmetry in our class of models is the
product of $R$ symmetry and $Aut(G)$.
This extension of ${\bf R}_i$ can break SUSY as well as the bulk gauge group $G$.
Depending on the discrete charge assignment,
the 4D $N=4$ SUSY can be broken down to $N= 0,1,2$.

If at least $N=1$ SUSY remains at 4D fixed points, the orbifold
conditions of the superfields $V$ and $\Sigma_i$ are given as
\begin{eqnarray}
V(x^\mu,\bar\omega \bar z_i,\omega z_i) &=& R_i[V(x^\mu,\bar z_i,z_i)], \\
\Sigma_1(x^\mu,\bar\omega \bar z_i,\omega z_i) &=& \bar \omega^{k_i} R_i[ \Sigma_1(x^\mu,\bar z_i,z_i) ], \\
\Sigma_2(x^\mu,\bar\omega \bar z_i,\omega z_i) &=& \bar \omega^{l_i} R_i[ \Sigma_2(x^\mu,\bar z_i,z_i) ], \\
\Sigma_3(x^\mu,\bar\omega \bar z_i,\omega z_i) &=& \bar \omega^{m_i} R_i[ \Sigma_3(x^\mu,\bar z_i,z_i) ],
\end{eqnarray}
where $R_i$ acts on the gauge algebra.
Since there is a higher dimensional version of trilinear gauge interaction term
%${\rm tr}\,\Phi_1 [\Phi_2, \Phi_3]$,
in Lagrangian,
$k_i+l_i+m_i  \equiv 0$ (mod $n_i$) needs to be satisfied.
Also, we need $k_1 = 1$, $l_2 =1$ and $m_3 = 1$ to make the lagrangian invariant.
Therefore, one of $(k_i,l_i,m_i)$ has to be 1.
From a geometrical consequence, $n_i$ has to be 2,3,4,6. Possible
combinations of $(k_i,l_i,m_i)$ are then the followings up to
permutation:
\begin{eqnarray}
n_i =2,  && (k_i,l_i,m_i) = (0,1,1), \\
n_i =3,  && (k_i,l_i,m_i) = (0,1,2), (1,1,1), \\
n_i =4,  && (k_i,l_i,m_i) = (0,1,3), (1,1,2), \\
n_i =6,  && (k_i,l_i,m_i) = (0,1,5), (1,1,4), (1,2,3).
\end{eqnarray}
For example, let us consider the case where the orbifolds for $z_2,
z_3$ are trivial (we may call the case $n_2,n_3 = 1$.). If this is
the case to build a model, we may consider 6D model instead of 10D
model. In this case, $(k_1,l_1,m_1) = (1,l,m)$, and $1+l+m \equiv 0$
(mod $n_1$). If $l = 0$, $N=2$ SUSY remains on the orbifold fixed
points.
In general, if one of $(k_i,l_i,m_i)$ is zero, $N=2$ remains on the fixed points from the action ${\bf R}_i$.
However, if the other torus orbifold condition is non-trivial, the SUSY can be broken down to $N=1$
for the 4D zero modes.
In the examples we will see later,
one can break down SUSY down to $N=1$ as well as the bulk gauge group even if $n_3=1$.
Therefore, the models can be constructed even in 8D.
In that case, the scalar component of $\Sigma_3$
is not an extra dimensional gauge bosons.

The scalar components of the superfields $\Sigma_i$ are higher
dimensional gauge fields. As a consequence, the bulk gauge
interaction includes the term $f_{abc} \,\Sigma_1^a \Sigma_2^b
\Sigma_3^c$ in the superpotential in 4D, where $f_{abc}$ is a
structure constant of the gauge group. If the matter and Higgs
fields originate from the gauge multiplet, the bulk gauge
interaction can contain the Yukawa coupling to generate the fermion
masses by Higgs mechanism. Actually, when we use the conventional
normalization of the gauge coupling, the bulk Yukawa coupling
constant is the same as the gauge coupling $g$ in 4D. Therefore, it
is interesting to consider the possibility that the fermions (quarks
and leptons), the gauge bosons and the Higgs fields, all originate
as the zero modes from the same higher dimensional gauge multiplet.

\section{Charge assignments in Pati-Salam basis}

In this section, we will assign the discrete charges to the
decomposed representations under Pati-Salam symmetry,
which are contained in the adjoint representation
of $E_8$, $E_7$, and $SO(16)$.
Under the Pati-Salam branch, it is clear to see
all the chiral matter species in the adjoint representations
of bulk gauge symmetries.

It is well known that $E_8$ has a maximal subgroup $SO(16)$,
and thus it can contain $SO(10)$ gauge symmetry with flavor symmetry $SO(6) \simeq SU(4)$.
The adjoint representation $\bf 248$ is decomposed under $SO(10)\times SU(4)$
as
\begin{equation}
{\bf 248} = ({\bf 45},{\bf 1}) + ({\bf 16},{\bf 4}) + (\overline{\bf 16},\overline{\bf 4})
+ ({\bf 10},{\bf 6}) + ({\bf 1},{\bf 15}).
\end{equation}
The $SO(10)$ has a Pati-Salam subgroup $G_{\rm PS} = SU(4) \times SU(2) \times SU(2)$,
and it is useful to describe the $SO(10)$ algebra in the Pati-Salam basis.
Here, we assign the $\mathbb{Z}_n$ charges for the decomposed fields under the Pati-Salam symmetry
to break the bulk gauge symmetry $E_8$ down to the Pati-Salam symmetry.

The $SO(10)$ adjoint field contains the following decomposed representations and
the discrete charges are assigned as follows:
\begin{equation}
\begin{array}{|c|c|c|c|}
\hline
({\bf 15},{\bf 1},{\bf 1}) & ({\bf 1},{\bf 3},{\bf 1}) & ({\bf 1},{\bf 1},{\bf 3})
& ({\bf 6},{\bf 2},{\bf 2}) \\
\hline \hline
0 & 0 & 0 & e \\
\hline
\end{array}
\end{equation}

The spinor representation $\bf 16$ contains four flavors of matter $L_i : ({\bf 4},{\bf 2},{\bf 1})$,
and $\bar R_i : (\bar{\bf 4},{\bf 1},{\bf 2})$.
\begin{equation}
\begin{array}{|c|c|c|c|c|c|c|c|}
\hline
L_1 & L_2 & L_3 & L_4 & \bar R_1 & \bar R_2 & \bar R_3 & \bar R_4 \\
\hline \hline
x & y & z & w & e+x & e+y & e+z & e+w \\
\hline
\end{array}
\end{equation}

The spinor representation $\overline{\bf 16}$ contains four flavors of anti-matter
$\bar L^i : (\bar{\bf 4},{\bf 2},{\bf 1})$,
and $R^i : ({\bf 4},{\bf 1},{\bf 2})$.
\begin{equation}
\begin{array}{|c|c|c|c|c|c|c|c|}
\hline
\bar L^1 & \bar L^2 & \bar L^3 & \bar L^4 & R^1 & R^2 & R^3 & R^4 \\
\hline \hline
-x & -y & -z & -w & e-x & e-y & e-z & e-w \\
\hline
\end{array}
\end{equation}

The vector representation $\bf 10$ includes $SU(4)$ sextet $C^{ij} : ({\bf 6},{\bf 1},{\bf 1})$,
and bidoublet representation $H^{ij} : ({\bf 1},{\bf 2},{\bf 2})$.
They are sextet under the flavor symmetry, and the flavor indices $ij$ are anti-symmetric.
\begin{equation}
\begin{array}{|c|c|}
\hline
 C^{ij}  &  H^{ij}  \\
\hline \hline
\left(\begin{array}{cccc}
{\mbox -}  & z+w & y+w & y+z \\
  &  {\mbox -}   & x+w & x+z \\
  &     &  {\mbox -}   & x+y \\
  &     &     & {\mbox -}
 \end{array} \right) &
\left(\begin{array}{cccc}
{\mbox -}  & e+z+w & e+y+w & e+y+z \\
  & {\mbox -}   & e+x+w & e+x+z \\
  &     &  {\mbox -}   & e+x+y \\
  &     &     & {\mbox -}
 \end{array} \right) \\
\hline
\end{array}
\end{equation}
Because $\epsilon_{ijkl} C^{ij} C^{kl}$ is a singlet, $C^{34}$ component is a conjugate of
$C^{12}$, for example.

The $SO(10)$ singlet $S_i{}^j$ is an adjoint under the flavor symmetry.
\begin{equation}
\begin{array}{|c|}
\hline
 S_{i}{}^j \\
\hline \hline
\left(\begin{array}{cccc}
 0 & x-y & x-z & x-w \\
-x+y  &  0   & y-z & y-w \\
-x+z  &  -y+z   &  0   & z-w \\
-x+w  &  -y+w   &  -z+w   & 0
 \end{array} \right) \\
\hline
\end{array}
\end{equation}

The bulk gauge interaction includes the Yukawa term $L_i \bar R_j H^{ij}$.

There are constraints for the discrete charges $x,y,z,w$ and $e$ due to the algebra:
\begin{equation}
x+y+z+w \equiv 0, \qquad 2e \equiv 0 \ \ (\mbox{mod } n).
\end{equation}
When $e\equiv 0$, at least $SO(10)$ symmetry remains
because the self-conjugate representation $({\bf 6},{\bf 2},{\bf 2})$
has a zero mode in vector multiplet.
Under these conditions, one can find that $T^2/\mathbb{Z}_3$
orbifold can not break $SO(10)$ symmetry for these Pati-Salam decomposition.

As is given, one $E_8$ adjoint includes 4 flavors
in $SO(10)_{\rm GUT} \times SU(4)_F$ branch.
One can also interpret the $SU(4)$ symmetry as $SU(4)_c$,
which includes color $SU(3)_c$ symmetry.
%there are 2 flavors in one adjoint.
In the former case, there is a self-conjugate representation $({\bf
6},{\bf 2},{\bf 2})$ under the Pati-Salam decomposition, and thus
with $T^2/\mathbb{Z}_3$ orbifold, it is impossible to break $SO(10)$
symmetry as we have noted.
%
%since the $Z_n$ charge of self conjugate must be $0$ or $n/2$.
%(When the charge is zero, higher symmetry remains.)
In the latter branch of $SU(4)_c \times SO(10)_w$, however, the self-conjugate representation is not included
under Pati-Salam symmetry
and thus $T^2/\mathbb{Z}_3$ orbifold can break
$SO(10)_w$ down to $SU(2)_L \times SU(2)_R \times U(1)^3$.
%Therefore, we can consider $T^2/Z_3$ orbifold in the case,
%\begin{equation}
%E_8 \to SU(4)_c \times SO(10)_w \to SU(4)_c \times SU(2)_L \times SU(2)_R \times U(1)^3.
%\end{equation}
%This is also true if we use $SU(9)$ branch.
Actually, this is related to the fact that
$E_8$ has two different $SU(8)\times U(1)$ subgroups.
One is obtained from $E_7$ branch, $E_8 \to E_7 \times U(1) \to SU(8) \times U(1)$:
\begin{equation}
{\bf 248} = {\bf 63}_0 + {\bf 28}_2 + \overline{\bf 28}_{-2} + {\bf 1}_{0} +
{\bf 1}_{-4} + {\bf 28}_{-2} + {\bf 70}_0 + \overline{\bf 28}_{2} + {\bf 1}_{4}.
\end{equation}
%This is obtained from $E_7$ branch.
The other is obtained from $SU(9)$ branch $E_8 \to SU(9) \to SU(8) \times U(1)$:
\begin{equation}
{\bf 248} = {\bf 63}_0 + {\bf 28}_2 + \overline{\bf 28}_{-2} + {\bf 1}_{0} +
{\bf 8}_{-3} + {\bf 56}_{-1} + \overline{\bf 56}_{1} + \overline{\bf 8}_{3}.
\end{equation}
%This is obtained from $SU(9)$ branch.
The former one includes a self-conjugate representation $\bf 70$,
four-rank anti-symmetric tensor in $SU(8)$,
but the latter one does not have any self conjugate representations.
The $SU(8)$ adjoint $\bf 63$ includes one flavor of matter under the Pati-Salam decomposition,
and $\bf 28$ and $\overline{\bf 28}$ include one more flavor.
Therefore, in the latter branch, there are two flavors under the Pati-Salam branch in the
adjoint representation.
If we use the $SU(9)$ branch, it is easy to obtain the charge assignment
for the Pati-Salam decomposed fields.
The adjoint representation is decomposed as ${\bf 248} = {\bf 80}+ {\bf 84} + \overline{\bf 84}$
under $SU(9)$,
where $\bf 80$ is an adjoint under $SU(9)$ and $\bf 84$ is a three-anti-symmetric tensor.
When we assign the $\mathbb{Z}_{n}$ charge as
\begin{equation}
\begin{array}{|c|c|c|}\hline
{\bf 80} & {\bf 84} & \overline{\bf 84} \\ \hline \hline
  0      &   a      &  -a \\ \hline
\end{array}
\end{equation}
$E_8$ is broken down to $SU(9)$.
Under the $SU(9)$ space, acting the unitary rotation matrix
\begin{equation}
R = {\rm diag} \, ( 1,1,1,1,\omega^b, \omega^b, \omega^c, \omega^c, \omega^d),
\end{equation}
we obtain the charge assignments for the component representations in the Pati-Salam branch.
For example, the left-handed matter components $({\bf 4},{\bf 2},{\bf 1})$
are in the $\bf 80$ and $\bf 84$, and their
charges are found to be $-b$ and $a+b+d$.

We can also obtain $\mathbb{Z}_n$ charges under the Pati-Salam decomposition
using the breaking chain $E_7 \to SO(12)\times SU(2) \to SO(10)\times SU(2)_F \times U(1)$.
In this case, there are two flavors in the $E_7$ adjoint representation:
\begin{equation}
{\bf 133} = ({\bf 45},{\bf 1})_0 + ({\bf 10},{\bf 1})_{-2} + ({\bf 10},{\bf 1})_2 +({\bf 1},{\bf 1})_0
+ ({\bf 16},{\bf 2})_1+ (\overline{\bf 16},{\bf 2})_{-1} + ({\bf 1},{\bf 3})_0.
\end{equation}

The $SO(10)$ adjoint field contains the fields as follows:
\begin{equation}
\begin{array}{|c|c|c|c|}
\hline
({\bf 15},{\bf 1},{\bf 1}) & ({\bf 1},{\bf 3},{\bf 1}) & ({\bf 1},{\bf 1},{\bf 3})
& ({\bf 6},{\bf 2},{\bf 2}) \\
\hline \hline
0 & 0 & 0 & e \\
\hline
\end{array}
\end{equation}

The spinor representation $\bf 16$ contains four flavors of matter $L_i : ({\bf 4},{\bf 2},{\bf 1})$,
and $\bar R_i : (\bar{\bf 4},{\bf 1},{\bf 2})$,
and
the spinor representation $\overline{\bf 16}$ contains four flavors of anti-matter
$\bar L^i : (\bar{\bf 4},{\bf 2},{\bf 1})$,
and $R^i : ({\bf 4},{\bf 1},{\bf 2})$.
\begin{equation}
\begin{array}{|c|c|c|c||c|c|c|c|}
\hline
L_1 & L_2 & \bar R_1 & \bar R_2 & \bar L^1 & \bar L^2  & R^1 & R^2 \\
\hline \hline
x & y & e+x & e+y & -x & -y  & e-x & e-y\\
\hline
\end{array}
\end{equation}

%The spinor representation $\overline{\bf 16}$ contains four flavors of anti-matter
%$\bar L_i : (\bar{\bf 4},{\bf 2},{\bf 1})$,
%and $R_i : ({\bf 4},{\bf 1},{\bf 2})$.
%\begin{equation}
%\begin{array}{|c|c|c|c|}
%\hline
%\bar L^1 & \bar L^2  & R^1 & R^2  \\
%\hline \hline
%-x & -y  & e-x & e-y \\
%\hline
%\end{array}
%\end{equation}

The vector representation $\bf 10$ includes $SU(4)$ sextet $C$ and $\bar C$ $: ({\bf 6},{\bf 1},{\bf 1})$,
and bidoublet representation $H$ and $\bar H$ $: ({\bf 1},{\bf 2},{\bf 2})$.
The fields $C$ and $\bar C$ have opposite $U(1)$ charges.
\begin{equation}
\begin{array}{|c|c|c|c|}
\hline
 C & \bar C  &  H & \bar H  \\
\hline \hline
-x-y & x+y & e-x-y & e+x+y \\
\hline
\end{array}
\end{equation}

The $SO(10)$ singlets $S_i{}^j$ are the adjoint under the flavor symmetry $SU(2)$.
\begin{equation}
\begin{array}{|c|}
\hline
 S_{i}{}^j \\
\hline \hline
\left(\begin{array}{cc}
 0 & x-y  \\
-x+y  &  0
 \end{array} \right) \\
\hline
\end{array}
\end{equation}

Similarly to the previous case, we need $2 e \equiv 0$.
The bulk interaction includes a term $\epsilon_{ij} L_i \bar R_j H$.

We also note that the adjoint representation of $SO(16)$
contains two families of matters in the Pati-Salam basis
using the chain $SO(16) \to SU(8) \times U(1)$ \cite{Gogoladze:2003yw}.
The adjoint representation of $SO(16)$
is decomposed as ${\bf 120} = {\bf 63}+ {\bf 28} + \overline{\bf 28}+ {\bf 1}$
under $SU(8) \times U(1)$,
where $\bf 63$ is a adjoint under SU(8) and $\bf 28$ is an anti-symmetric tensor.
The charge assignment to obtain the Pati-Salam branch
is similar to the case of the $SU(9)$ branch in $E_8$ we have seen.
When we assign the $\mathbb{Z}_{n}$ charge as
\begin{equation}
\begin{array}{|c|c|c|c|}\hline
{\bf 63} & {\bf 28} & \overline{\bf 28} & {\bf 1} \\ \hline \hline
  0      &   a      &  -a  & 0  \\ \hline
\end{array}
\label{so16-1}
\end{equation}
$SO(16)$ is broken down to $SU(8)\times U(1)$.
Under the $SU(8)$ space, acting the rotation matrix
\begin{equation}
R = {\rm diag} \, ( 1,1,1,1,\omega^b, \omega^b, \omega^c, \omega^c),
\label{so16-2}
\end{equation}
we obtain the charge assignments for the component representations in the Pati-Salam branch.
For example, the left-handed matter components $({\bf 4},{\bf 2},{\bf 1})$
are in the $\bf 63$ and $\bf 28$, and their
charges are $-b$ and $a+b$.

\section{Examples for three families in the bulk}

We will classify the cases where there are just three chiral flavors,
which are obtained from the zero modes of the bulk fields.
For the Pati-Salam decomposition, $E_8$ adjoint representation
contain four vector-like matter representations,
as it is written in the previous section.
Since there are three adjoint superfields, $\Sigma_i$,
there are totally 12 vector-like matters in the bulk.
By orbifold projection, we will extract only three flavors of chiral matter.

One can find that there are the following five ways to extract the three flavors.
The left-handed matter components $L_i : ({\bf 4},{\bf 2},{\bf 1})$ under Pati-Salam
symmetry have flavor index $i$ for the $SU(4)$ fundamental representation.
%which is obtained from the subgroup $SO(10)\times SU(4)_F \subset E_8$.
In the following,
we do not care about the permutation of the flavor indices,
as well as the permutation of $\Sigma_1$, $\Sigma_2$, and $\Sigma_3$.

\begin{enumerate}
    \item
    The three chiral superfields $\Sigma_1$, $\Sigma_2$, and $\Sigma_3$
    include $L_1$, $L_2$, and $L_3$, respectively.

    \item
    The three chiral superfields $\Sigma_1$, $\Sigma_2$, and $\Sigma_3$
    include $L_1$, $L_1$, and $L_2$, respectively.

    \item
    The three chiral superfields $\Sigma_1$, $\Sigma_2$, and $\Sigma_3$
    include $L_1$, $L_1$, and $L_1$, respectively.

    \item
    Flavor $SU(2)$ symmetry remains in 4D, and $\Sigma_1$ and $\Sigma_2$
    include $(L_1,L_2)$ and $L_3$ respectively.

    \item
    Flavor $SU(3)$ symmetry remains in 4D, and $\Sigma_1$ includes
    $(L_1,L_2,L_3)$.

\end{enumerate}

In the case 2, we need only two flavors of matter in the adjoint representation,
and thus, the bulk symmetry can be $E_7$ or $SO(16)$.
In the case 3, only one flavor of matter is needed,
and so, the bulk symmetry can be $SU(8)$ \cite{Gogoladze:2007nb}.
Also, in the case 3, all three chiral superfields $\Sigma_i$ must have
the same $\mathbb{Z}_{n_i}$ charge assignment $k_i=l_i=m_i$.
Therefore, $n_i$ has to be 3 in this case.

In the following, we will give examples of charge assignment
in each case.

\subsection{Case 1}

The three chiral superfields $\Sigma_1$, $\Sigma_2$, and $\Sigma_3$
include $L_1$, $L_2$, and $L_3$ components, respectively.
To do this, it is necessary that
\begin{equation}
-k_i+x =0, \quad -l_i+y=0, \quad -m_i+z = 0.
\end{equation}
The equivalence symbol with mod $n$ is omitted in the following.
Since $k_i+l_i+m_i=0$ and $x+y+z+w = 0$, we need $w=0$.
As a consequence, the corresponding component in the vector field
must have a zero mode, and thus the 4D gauge symmetry is always higher than Pati-Salam
symmetry in this case.
If only $L_4,\bar L_4$ components remain massless in the vector multiplet,
the 4D gauge symmetry will be $SU(6)\times SU(2) \times U(1)^2$.
When $R_4$ and $\bar R_4$ remain mass less in addition to them ($w=e=0$),
$E_6 \times U(1)^2$ symmetry remains in 4D.

\bigskip

{\bf Example 1}:
Consider the orbifold $T^2/\mathbb{Z}_6 \times T^2/\mathbb{Z}_3$
($n_1 =6$, $n_2 =3$).

The charge assignments are
\begin{eqnarray}
&&(x,y,z,w)_{{Z}_6} = (1,2,3,0), \quad (x,y,z,w)_{{Z}_3} = (1,1,1,0), \quad
e_{{Z}_6} = e_{{Z}_3} = 0, \\
&&(k_1,l_1,m_1) = (1,2,3), \quad (k_2,l_2,m_2)= (1,1,1).
\end{eqnarray}

Under these assignments, one can find that the 4D gauge symmetry is
$E_6 \times U(1)^2$.
Three chiral fields ${\bf 27}_i$ are zero modes.
Since $E_6$ symmetry remains in this example, it is better to see the
chain $E_8 \to E_6\times SU(3)$ and the decomposition of the adjoint field:
\begin{equation}
{\bf 248} = ({\bf 78},{\bf 1})+({\bf 27},{\bf 3})+(\overline{\bf 27},\overline{\bf 3})+({\bf 1},{\bf 8}).
\label{E_6 branch}
\end{equation}
In fact, under the above charge assignments, the decomposed fields are rearranged
as in the $E_6$ representation, and the $SU(3)$ symmetry is broken down to $U(1)^2$.
The bulk interaction includes the Yukawa term
\begin{equation}
\mathbf{27}_1 \cdot \mathbf{27}_2 \cdot \mathbf{27}_3.
\end{equation}
When the bidoublet components $H^{12},H^{23},H^{31}$ (which are also unified in ${\bf 27}_i$) get VEVs,
the fermions acquire masses.
Since $\overline{\bf 27}$ from the product of ${\bf 27}\times {\bf 27}$ is symmetric under $E_6$ algebra,
%and they are also totally anti-symmetric under the indices $SU(3)$ symmetry,
the mass matrix of the fermion is symmetric.
This is because $\bf 27$ includes both left- and right-handed matters.
Note that the indices $i$ of ${\bf 27}_i$ are the flavor indices,
and
the bulk interaction is given as tr $[\Sigma_1, \Sigma_2] \Sigma_3$.
As a result, the above Yukawa term is included in the bulk interaction,
even though the flavor indices are totally anti-symmetric.

Since the diagonal entries are all zero in the symmetric mass matrix, the eigenvalues of the mass matrix
cannot be hierarchical (two of the eigenvalues are nearly degenerate,
or all three are of the same order).
Therefore, to obtain the realistic fermion masses, we need an assist from the brane-localized
terms.

\bigskip

{\bf Example 2}:
Next, let us consider the case where the 4D gauge symmetry is
$SU(6)\times SU(2)_L \times U(1)^2$.
This is obtained when $R_4,\bar R_4$ components in the vector superfield
remain zero modes ($e+w=0$).
We consider
the orbifold $T^2/\mathbb{Z}_4 \times T^2/\mathbb{Z}_3$
($n_1 =4$, $n_2 =3$).
The charge assignments are
\begin{eqnarray}
&&(x,y,z,w)_{{Z}_4} = (1,3,2,2), \quad (x,y,z,w)_{{Z}_3} = (1,1,1,0), \quad
e_{{Z}_4} = 2, \quad e_{{Z}_3} = 0, \\
&&(k_1,l_1,m_1) = (1,0,3), \quad (k_2,l_2,m_2)= (1,1,1).
\end{eqnarray}
Then we obtain the zero modes in the chiral superfields as
\begin{equation}
\begin{array}{c|c}
\Sigma_1  & ({\bf 6},{\bf 2})_1, (\overline{\bf 15},{\bf 1})_2 \\
\Sigma_2  & (\overline{\bf 15},{\bf 1})_3 \\
\Sigma_3  & ({\bf 6},{\bf 2})_2, (\overline{\bf 15},{\bf 1})_1
\end{array}
\end{equation}
The right-handed fermions are includes in the representation $(\overline{\bf 15},{\bf 1})$,
and the left-handed fermions are included in $({\bf 6},{\bf 2})$.
So, there are two left-handed fermions and three right-handed fermions are
the zero modes of the bulk fields in this example.
One left-handed fermions must be a brane-localized field.
The bulk interaction includes the Yukawa term
\begin{equation}
({\bf 6},{\bf 2})_1 ({\bf 6},{\bf 2})_2 (\overline{\bf 15},{\bf 1})_3
+(\overline{\bf 15},{\bf 1})_2 (\overline{\bf 15},{\bf 1})_3 (\overline{\bf 15},{\bf 1})_1.
\end{equation}
The Higgs bidoublet components are also in the $({\bf 6},{\bf 2})_1$, and $({\bf 6},{\bf 2})_2$.
In this example, the bulk Yukawa mass matrix for fermions is rank one, and only one of the eigenstate is massive
from the bulk interaction.

\subsection{Case 2}

The three chiral superfields $\Sigma_1$, $\Sigma_2$, and $\Sigma_3$
include $L_1$, $L_1$, and $L_2$ components, respectively.
To do this, it is necessary that
\begin{equation}
-k_i+x =0, \quad -l_i+x=0, \quad -m_i+y = 0.
\end{equation}

In this case, we can extract the zero modes of three chiral families,
when (at least) two families are included in the adjoint representation.
So, let us consider the example by using a model with $E_7$ bulk symmetry.

\bigskip

{\bf Example 3}:
We consider
the orbifold $T^2/\mathbb{Z}_6 \times T^2/\mathbb{Z}_3$
($n_1 =6$, $n_2 =3$).
The charge assignments for the decomposed fields from $E_7$ are
\begin{eqnarray}
&&(x,y)_{{Z}_6} = (1,4), \quad (x,y)_{{Z}_3} = (1,1), \quad
e_{{Z}_6} = 3, \quad e_{{Z}_3} = 0, \\
&&
(k_1,l_1,m_1) = (1,1,4), \quad (k_2,l_2,m_2)= (1,1,1).
\end{eqnarray}

Then, the 4D symmetry is $G_{\rm PS} \times U(1)^2$, and the zero modes are
\begin{equation}
\begin{array}{c|c}
\Sigma_1 &  L_1, \bar R_2^\prime, C^{\prime}  \\
\Sigma_2 &  L_1^\prime, \bar R_2, C\\
\Sigma_3 &  L_2, \bar R_1, H
\end{array}
\end{equation}
Since the $L_1$ components in the superfields $\Sigma_1$ and $\Sigma_2$ are different fields,
we denote them $L_1$ and $L_1^\prime$.
We obtain the massless modes for the three families and one Higgs bidoublet,
and two sextet fields.

The bulk interaction is
\begin{equation}
(L_1 \bar R_2 + L_1^\prime \bar R_2^\prime) H
+ C (L_1 L_2 - \bar R_1 \bar R_2^\prime)
+ C^{\prime} (L_1^\prime L_2 + \bar R_1 \bar R_2).
\end{equation}
One can find that $(L_1, L_1^\prime)$, $(\bar R_2, \bar R_2^\prime)$, and
$(C,C^\prime)$ form doublets under global $SU(2)$ symmetry,
which originates from an $R$-symmetry for the gauge multiplet in the bulk.
When the Higgs bidoublet $H$ gets VEV,
the fermions acquire masses.
The two eigenvalues are degenerate and one eigenstate is massless.

\subsection{Case 3}

The three chiral superfields $\Sigma_1$, $\Sigma_2$, and $\Sigma_3$
include $L_1$, $L_1$, and $L_1$ components, respectively.
To do this, it is necessary that
\begin{equation}
-k_i+x =0, \quad -l_i+x=0, \quad -m_i+x = 0.
\end{equation}
Since $k_i=l_i=m_i$ needs to be satisfied,
only $\mathbb{Z}_3$ orbifold is possible in this case.

The model can be constructed if (at least) one family
is contained in the adjoint representation,
and thus the bulk gauge symmetry can be $SU(8)$.
We can construct a model by 6D $T^2/\mathbb{Z}_3$ orbifold with
$SU(8)$ bulk gauge symmetry \cite{Gogoladze:2007nb}.

When we consider 8D $T^2/\mathbb{Z}_3 \times T^2/\mathbb{Z}_3^\prime$ orbifold,
one can obtain the model with $SO(16)$ and $E_8$ bulk gauge symmetries.
Here, we give a solution for the case of bulk $SO(16)$ symmetry.
The charge assignments for the decomposed fields from $SO(16)$, Eqs. (\ref{so16-1},\ref{so16-2}), are
\begin{eqnarray}
&&(a,b,c)_{{Z}_3} = (1,2,1), \quad (a,b,c)_{{Z}_3^\prime} = (2,2,1), \\
 &&(k_1,l_1,m_1) = (1,1,1), \quad
(k_2,l_2,m_2)= (1,1,1).
\end{eqnarray}

%---------------------------------------------------------------------------
%NOTE: SHOULD a, b, c be x, y, z??
%-------------------------------------------------------------------------

In these solutions,
there are three chiral families as well as three Higgs bidoublet fields
as the zero modes from the bulk gauge multiplet.
The bulk Yukawa coupling includes
\begin{equation}
\epsilon_{ijk} L_i \bar R_j H_k,
\end{equation}
where $\epsilon_{ijk}$ is a totally anti-symmetric tensor.
Note that $i,j$ and $k$ are not the indices from the $SU(4)$ subgroup in $E_8$,
but they are the indices originate from $\Sigma_1$, $\Sigma_2$, and $\Sigma_3$.
The Yukawa couplings for the zero modes have a global $SU(3)$ symmetry.
%which originates from an $R$-symmetry in the bulk.

Since the Yukawa matrix is anti-symmetric,
two eigenvalues of fermion masses are degenerate
and one eigenstate is massless
when the Higgs fields $H_i$ get VEVs.
The detail is given in the Ref.\cite{Gogoladze:2007nb}.

\subsection{Case 4}

Flavor $SU(2)$ gauge symmetry remains in 4D, and $\Sigma_1$ and $\Sigma_2$
include $(L_1,L_2)$ and $L_3$ respectively.
The field $(L_1,L_2)$ in $\Sigma_1$ forms a doublet under the flavor $SU(2)$ symmetry.
To satisfy this choice, it is necessary that
\begin{equation}
-k_i+x =0, \quad x=y, \quad -l_i+z=0.
\end{equation}

\bigskip

{\bf Example 4}:
We consider
the orbifold $T^2/\mathbb{Z}_6 \times T^2/\mathbb{Z}_4$
($n_1 =6$, $n_2 =4$).
The charge assignments for the decomposed fields are
\begin{eqnarray}
&&(x,y,z,w)_{{Z}_6} = (1,1,4,0), \quad (x,y,z,w)_{{Z}_4} = (1,1,1,1), \quad
e_{{Z}_6} = 3, \quad e_{{Z}_4} = 0, \\
&&(k_1,l_1,m_1) = (1,4,1), \quad (k_2,l_2,m_2)= (1,1,2).
\end{eqnarray}

Then, the 4D symmetry is $G_{\rm PS} \times SU(2) \times U(1)^2$, and the zero modes are
\begin{equation}
\begin{array}{c|c}
\Sigma_1 &  L_1, L_2, \bar R_3  \\
\Sigma_2 &  L_3, \bar R_1, \bar R_2 \\
\Sigma_3 &  C^{23}, C^{13}, H^{12}
\end{array}
\end{equation}
The fields $(L_1,L_2)$, $(\bar R_1,\bar R_2)$, and $(C^{13},C^{23})$ form
$SU(2)$ doublets.
The bulk interaction is
\begin{equation}
(L_1 \bar R_2 + L_2 \bar R_1) H^{12} + L_3 (L_1 C^{13} + L_2 C^{23}) +
\bar R_3 (-\bar R_1 C^{13}+ \bar R_2 C^{23}).
\end{equation}

When Higgs bidoublet $H^{12}$ get a VEV,
the fermions acquire masses.
The two eigenvalues are degenerate, and one eigenstate (actually $L_3$ and $\bar R_3$) remain massless.

\bigskip

{\bf Example 5}:
We consider
the orbifold $T^2/\mathbb{Z}_6 \times T^2/\mathbb{Z}_4$
($n_1 =6$, $n_2 =4$).
The charge assignments for the decomposed fields are
\begin{eqnarray}
&&(x,y,z,w)_{{Z}_6} = (5,5,0,2), \quad (x,y,z,w)_{{Z}_4} = (1,1,1,1), \quad
e_{{Z}_6} = 3, \quad e_{{Z}_4} = 0, \\
&& (k_1,l_1,m_1) = (1,0,5), \quad (k_2,l_2,m_2)= (2,1,1).
\end{eqnarray}

Then, the 4D symmetry is $G_{\rm PS} \times SU(2) \times U(1)^2$, and the zero modes are
\begin{equation}
\begin{array}{c|c}
\Sigma_1 &  C^{23}, C^{13}, H^{34} \\
\Sigma_2 &  L_3 \\
\Sigma_3 &  L_1, L_2, \bar R_4
\end{array}
\end{equation}
In this example, three left-handed matters are zero modes,
and one right-handed matter is zero modes.
The fields $(L_1,L_2)$ and $(C^{13},C^{23})$ are $SU(2)$ doublets.

The bulk interaction is
\begin{equation}
L_3 \bar R_4 H^{34} + L_3 (L_1 C^{13} + L_2 C^{23}).
\end{equation}
When the Higgs bidoublet $H^{34}$ get VEV,
one eigenstate of fermion becomes massive.
The other two left-handed fields $(L_1,L_2)$
which is a flavor $SU(2)$ doublet, remain massless.
They can become massive by brane localized terms.
When we identify $(L_1,L_2)$ as the first and second generations,
one can construct a flavor $SU(2)$ model
to obtain a hierarchical pattern.

\subsection{Case 5}

Flavor $SU(3)$ gauge symmetry remains in 4D, and $\Sigma_1$ includes
$(L_1,L_2,L_3)$.
To satisfy this choice, it is necessary that
\begin{equation}
-k_i+x =0, \quad x=y=z.
\end{equation}

\bigskip

{\bf Example 6}:
We consider the
the orbifold $T^2/\mathbb{Z}_4 \times T^2/\mathbb{Z}_3$
($n_1 =4$, $n_2 =3$).
The charge assignments for the decomposed fields are
\begin{eqnarray}
&&(x,y,z,w)_{{Z}_4} = (1,1,1,1), \quad (x,y,z,w)_{{Z}_3} = (1,1,1,0), \quad
e_{{Z}_4} = 2, \quad e_{{Z}_3} = 0, \\
&&(k_1,l_1,m_1) = (1,0,3), \quad (k_2,l_2,m_2)= (1,1,1).
\end{eqnarray}
Then, the 4D symmetry is $G_{\rm PS} \times SU(3) \times U(1)$, and the zero modes are
\begin{equation}
\begin{array}{c|c}
\Sigma_1 &  L_1,L_2,L_3 \\
\Sigma_2 &  H^{12},H^{13},H^{23},S_1{}^4,S_2{}^4,S_3{}^4 \\
\Sigma_3 &  \bar R_1, \bar R_2, \bar R_3
\end{array}
\end{equation}
In this example, three families of both left- and right-handed matters are zero modes,
and $L_i$, $\bar R_i$, $H^{ij}(=-H^{ji})$ are triplets under the flavor $SU(3)$ symmetry.
The PS singlets $S_i{}^4$ also form a triplet under $SU(3)$.

The bulk interaction is
\begin{equation}
\sum_{i,j=1,2,3} L_i \bar R_j H^{ij},
\end{equation}
and the Yukawa matrix is anti-symmetric.

When the Higgs fields get VEVs, the flavor symmetry $SU(3)$ is broken down to $SU(2)$,
and the two eigenvalues are degenerate, and one eigenstate is massless.
(Surely, in a phenomenological model construction, the flavor symmetry must be broken at a higher scale
by SM singlet VEVs.)
The bulk Yukawa interaction is same as the case 3,
where the $SU(3)$ symmetry in the bulk interaction is not gauged,
while in this example, the flavor $SU(3)$ symmetry is a gauge symmetry.

\bigskip

{\bf Example 7}:
We consider
the orbifold $T^2/\mathbb{Z}_6 \times T^2/\mathbb{Z}_4$
($n_1 =6$, $n_2 =4$).
The charge assignments for the decomposed fields are
\begin{eqnarray}
&&(x,y,z,w)_{{Z}_6} = (2,2,2,0), \quad (x,y,z,w)_{{Z}_4} = (1,1,1,1), \quad
e_{{Z}_6} = 3, \quad e_{{Z}_4} = 0, \\
&&(k_1,l_1,m_1) = (1,2,3), \quad (k_2,l_2,m_2)= (2,1,1).
\end{eqnarray}
Then, the 4D symmetry is $G_{\rm PS} \times SU(3) \times U(1)$, and the zero modes are
\begin{equation}
\begin{array}{c|c}
\Sigma_1 &  H^{14},H^{24},H^{34} \\
\Sigma_2 &  L_1,L_2,L_3 \\
\Sigma_3 &  \bar R_4
\end{array}
\end{equation}
In this example, three families of left-handed matters and one right-handed matter are zero modes.
The fields $L_i$, $H^{i4}$ are fundamental and anti-fundamental representation
 under the flavor $SU(3)$ symmetry.

The bulk interaction is
\begin{equation}
\sum_{i=1,2,3} L_i \bar R_4 H^{i4}.
\end{equation}

When the Higgs fields $H^{i4}$ get VEVs,
the flavor $SU(3)$ symmetry is broken down to $SU(2)$ symmetry.
Only one of the eigenstate is massive,
two other families can become massive by introducing the brane-localized interactions.
In this situation, it is easy to construct a model with hierarchical fermion masses
using the flavor symmetry \cite{Kitano:2000xk}.

\section{Comments to build a model}

As one can see, in the cases 1,2, and 4, the $SU(4)_c$ sextet fields
are also the zero modes from the bulk fields. In the case 1, one can
find that the $SU(4)_c$ sextet components are embedded in ${\bf 27}$
in Example 1 and $(\overline{\bf 15},{\bf 1})$ in Example 2. The
reason for this is as follows: In these cases, we need to satisfy
\begin{equation}
-k_i + x = 0, \quad -l_i + y = 0,
\end{equation}
up to the permutation of $(k_i,l_i,m_i)$ and $(x,y,z,w)$.
Because of a relation $k_i + l_i+ m_i = 0$, we need $x+y+m_i = 0$.
Since $-x-y$ is a charge of a $SU(4)$ sextet,
the corresponding sextet must be a zero mode.

If one thinks that the sextet field is unwanted,
one needs to choose case 3 or  5.
Actually, the sextet fields contain
colored Higgs fields which may generate a dangerous dimension 5 operator for proton decay.
So, if we add a brane-localized term,
a dangerous proton decay operator may be introduced.
However, the sextet fields have extra $U(1)$ charges,
and one can forbid the dangerous terms
if we do not introduce random PS singlet fields
which acquire VEV to break the extra $U(1)$ symmetries.
Furthermore,
since the representation of the (anti-fundamental) colored-Higgs fields
is same as the right-handed down-type quarks,
one can interpret that
the $SU(4)_c$ sextet contains the right-handed quarks,
and it can be available to break the strange quark and muon mass unification
(to obtain so-called Georgi-Jarskog relation).
It can be realized by mixing the fields with the right-handed quark in
$\bar R : (\bar{\bf 4},{\bf 1},{\bf 2})$.
The mixing terms will be introduced as a brane-localized term.
Or, one of $\bar R$ can be identified a Higgs field to break Pati-Salam symmetry
broken down to SM gauge symmetry.
In that case, the bulk term $C \bar R \bar R^\prime$ make $({\bf 3},{\bf 1})_{-1/3}$
component in $C$ massive, and the down-type quark component in $C$ remain massless.

To construct a model, we need to care about the brane-localized
gauge anomalies \cite{Scrucca:2004jn}. In general, the set of 4D
zero modes from the bulk chiral superfields causes the gauge
anomalies, and one has to introduce brane-localized fields at each
4D fixed point to cancel the anomalies. To cancel them, one has to
introduce the fields not to increase (or decrease) the number of
chiral families since our aim is to obtain the three chiral families
from the bulk. The introduction of the brane fields may be
complicated, but since it is not an essential part, we do not focus
it in this paper.

The main focus in this paper is the Yukawa coupling structure
originating from the bulk gauge interaction.
%
%The Yukawa couplings originate from the bulk gauge interaction.
%However, since the bulk interaction term is in the trilinear form
%$\Sigma_1 \Sigma_2, \Sigma_3$,
%the Yukawa matrix from the bulk interaction cannot
%have diagonal elements.
%As a consequence, the Yukawa matrix
%can not have three hierarchical eigenvalues ($m_1 \ll m_2 \ll m_3$).
%To obtain the realistic hierarchical fermion masses,
%we surely need an assist from the brane-localized terms.
%
The Yukawa matrix originating from the bulk interaction is
classified for the following three situations as we have seen in the
Examples in the previous section:

\medskip

(a) All three families can be massive.

(b) The two eigenvalues are degenerate, and one of the eigenstate is massless.

(c) One of the families is massive, and other two are massless.

\medskip

The situation (a) is a special case among the five cases in the previous section.
This is obtained from the case 1 in which the Yukawa interaction does not have
any global or gauged non-Abelian flavor symmetry.
In this case, however, if we see the Pati-Salam branch,
the 4D gauge symmetry is always larger than the PS symmetry,
and to obtain the situation (a), $E_6$ symmetry remains in the 4D fixed points.
In that case, the Yukawa matrix is symmetric under the flavor indices
and the diagonal elements are all zero.
As a consequence, the Yukawa matrix
can not have three hierarchical eigenvalues (namely, $m_1 \ll m_2 \ll m_3$).
To obtain realistic hierarchical fermion masses,
we surely need an assist from the brane-localized terms.

In the situation (b),
the Yukawa matrix is an anti-symmetric mass matrix
due to global or local $SU(3)$ symmetry,
or there is a global or local $SU(2)$ flavor symmetry in the bulk interaction.
This case is interesting to explain why the first generation of fermions
has tiny mass as is pointed out in Ref.\cite{Gogoladze:2007nb}.
To resolve the degeneracy between second and third generation,
we need to introduce the brane-localized term,
which is also studied in the Ref.\cite{Gogoladze:2007nb}.

In the situation (c),
only one family becomes massive because three left-handed matters are in the bulk,
but only one of the right-handed matters is in the bulk
as in the Examples 5 and 7.
This situation provides a good example to explain why only third generation is
heavy and their Yukawa coupling constant can be unified to the gauge coupling.
It is interesting that the flavor symmetry remains in 4D in the Examples
to construct a flavor model which can explain the hierarchy between first and second generation.

Among the three situations above, the situation (a) is special,
and it is possible to have a different fermion mass hierarchy,
if we see another branch of bulk gauge symmetry breaking.
Another typical branch which include SM gauge symmetry is
a trinification symmetry $G_{333} = SU(3)_c \times SU(3)_L \times SU(3)_R$.
Since $E_6$ has a maximal subgroup $G_{333}$,
it is possible to break the $E_8$ symmetry down to $E_6 \times SU(3) \to G_{333}\times U(1)^2$ by
orbifold \cite{Gogoladze:2003cp}.

\section{Trinification branch and SM decomposition}

\subsection{Charge assignments for trinification}

The $E_8$ group has a $E_6$ branch as in Eq.(\ref{E_6 branch}),
and $E_6$ has a trinification subgroup.
The adjoint ${\bf 78}$ and fundamental representation ${\bf 27}$ are
decomposed under $G_{333}$ as follows:
\begin{eqnarray}
{\bf 78} &=& ({\bf 8},{\bf 1},{\bf 1})+({\bf 1},{\bf 8},{\bf 1})+({\bf 1},{\bf 1},{\bf 8})+
({\bf 3},{\bf 3},{\bf 3})+(\bar{\bf 3},\bar{\bf 3},\bar{\bf 3}), \\
{\bf 27} &=& ({\bf 3},\bar{\bf 3},{\bf 1})+({\bf 1},{\bf 3},\bar{\bf 3})+(\bar{\bf 3},{\bf 1},{\bf 3}).
\end{eqnarray}
The representation ${\bf 27}$ includes $Q: ({\bf 3},\bar{\bf 3},{\bf 1})$,
$Q^c : ({\bf 1},{\bf 3},\bar{\bf 3})$, and $H: (\bar{\bf 3},{\bf 1},{\bf 3})$,
which include all the fermion species as well as Higgs representations
%which include left-handed quark doublet, right-handed quarks, and Higgs doublets,
under SM decomposition. %, respectively.
Note that the decomposed representation ${\bf 27}$ in $E_8$ has a $SU(3)$ flavor index.

The discrete charges are assigned to the decomposed representations as follows:
For the adjoint representations for $E_6$ and $SU(3)$, the charges are
\begin{equation}
\begin{array}{|c|c|c||c|}
\hline
({\bf 8},{\bf 1},{\bf 1})+({\bf 1},{\bf 8},{\bf 1})+({\bf 1},{\bf 1},{\bf 8})
 & ({\bf 3},{\bf 3},{\bf 3}) & (\bar{\bf 3},\bar{\bf 3},\bar{\bf 3}) & S_i{}^j \\
\hline \hline
0 & a & -a  & x_i - x_j \\
\hline
\end{array}
\end{equation}
where $S_i{}^j$ is a $SU(3)$ adjoint representation.

The matter and anti-matter representations are given as
\begin{equation}
\begin{array}{|c|c|c||c|c|c|}
\hline
 Q_i & Q_i^c & H_i & \bar Q^i & \bar Q^{ci} & \bar H^i \\
\hline \hline
x_i & a+x_i & -a+x_i & -x_i & -a -x_i & a-x_i  \\
\hline
\end{array}
\end{equation}

We need conditions $\sum x_i \equiv 0$, $3a \equiv 0$ (mod $n$).

\bigskip

{\bf Example 8}:
Let us obtain the case 1 solution considered in section 4
in $T^2/\mathbb{Z}_3 \times T^2/\mathbb{Z}_3^\prime$ orbifold.
The discrete charges are assigned as
\begin{eqnarray}
&&(x_1,x_2,x_3)_{{Z}_3} = (1,1,1), \quad (x_1,x_2,x_3)_{{Z}_3^\prime} = (0,1,2), \quad
a_{{Z}_3} = 0, \quad a_{{Z}_3^\prime} = 1, \\
&&(k_1,l_1,m_1) = (1,1,1), \quad (k_2,l_2,m_2)= (0,1,2).
\end{eqnarray}
Then, the 4D symmetry is $G_{333} \times U(1)^2$, and the zero modes are
\begin{equation}
\begin{array}{c|c}
\Sigma_1 &  Q_1, Q_3^c, H_2 \\
\Sigma_2 &  Q_2, Q_1^c, H_3 \\
\Sigma_3 &  Q_3, Q_2^c, H_1
\end{array}
\end{equation}
The bulk interaction includes
\begin{equation}
Q_1 Q_2^c H_3 + Q_2 Q_3^c H_1 + Q_3 Q_1^c H_2 + Q_1 Q_2 Q_3 + Q_1^c Q_2^c Q_3^c + H_1 H_2 H_3.
\end{equation}

Quarks $q,u^c,d^c$, leptons $\ell, e^c, \nu^c$ and the Higgs fields are
embedded in $Q$, $Q^c$ and $H$ as
\begin{equation}
Q = ( q , h^C ), \qquad Q^c = (u^c, d^c, \bar h^C), \qquad
H = \left( \begin{array}{ccc}
        h^u & h^d & L \\
        e^c & \nu^c & s
       \end{array}
    \right),
\end{equation}
where $h^u, h^d$ are Higgs doublets, $h^C$ and $\bar h^C$ are colored Higgs component,
and $s$ is a singlet component under SM.
Since they are embedded in $\bf 27$ representation of $E_6$,
the prediction of weak mixing angle is kept to be $\sin^2\theta_W = 3/8$
(as long as brane-localized contribution is suppressed).
Then, the bulk interaction includes the terms in terms of the SM decomposed fields
as
\begin{eqnarray}
Q_1 Q^c_2 H_3 &=& q_1 u^c_2 h^u_3 + q_1 d^c_2 h^d_3 + q_1 \bar h^C_2 \ell_3
+ h^C_1 d^c_2 e^c_3 + h^C_1 u^c_2 \nu^c_3 +h^C_1 \bar h^C_2 s_3, \\
Q_1 Q_2 Q_3 &=& q_1 q_2 h^C_3 + q_2 q_3 h^C_1 + q_3 q_1 h^C_2, \\
Q_1^c Q_2^c Q_3^c &=& u^c_1 d^c_2 \bar h^C_3 + u^c_2 d^c_3 \bar h^C_1 + u^c_3 d^c_1 h^C_2
+ d^c_1 u^c_2 \bar h^C_3 + d^c_2 u^c_3 \bar h^C_1 + d^c_3 u^c_1 h^C_2, \\
H_1 H_2 H_3 &=& \ell_1 e^c_2 h^d_3 + \ell_1 \nu^c_2 h^u_3 + s_1 h^u_2 h^d_3
+ (\mbox{permutation of indices}).
\end{eqnarray}

Because $Q^c_1$ component in $\Sigma_1$ is projected out for example,
the Yukawa matrix for quarks is no more symmetric.
Therefore, we do not have a restriction which we found in Example 1.
In a proper basis, Yukawa couplings for all three generations of quarks are unified to the gauge coupling.
When the ratios of VEVs of Higgs doublet fields $h^{u,d}_i$ are all free,
one can have any ratios of three-generation quark masses.
On the other hand,
the lepton Yukawa matrix is still symmetric because $H_i$ multiplet includes both
left- and right-handed leptons.
As it is mentioned, the symmetric matrix without diagonal elements can not have
three hierarchical eigenvalues.
%Since the bulk Yukawa matrix has no diagonal elements,
%two of lepton mass eigenvalues are degenerate
%when one of eigenvalues are chosen to be small.

It is preferable that
only one of the linear combination of Higgs doublet fields is light to construct
a low energy model
in order to avoid too large flavor changing neutral currents.
The gauge coupling unification also prefers the situation.
The bulk fields $H_i$ contain the standard model singlets $s_i$ as well as $SU(2)_L$ doublets
$h_i^u$, $h_i^d$,
%when the representation is decomposed in $G_{\rm SM}$.
and the bulk interaction $H_1 H_2 H_3$ includes the doublet Higgs mass terms $m_{ij} h_i^u h^d_j$
when $s_i$'s acquire VEVs,
\begin{equation}
m_{ij} = \left( \begin{array}{ccc}
    0 & s_3 & s_2 \\
    s_3 & 0 & s_1 \\
    s_2 & s_1 & 0
    \end{array} \right).
\end{equation}
The mass matrix $m_{ij}$
is symmetric and there is no diagonal elements since it comes from the bulk interaction and
$H_i$ multiplet includes both $h^u_i$ and $h^d_i$.
When the VEVs of singlets are hierarchical ($s_3 \ll s_2 \ll s_1$, for example),
only one of the linear combination of Higgs double fields is light,
and the Higgs mixings are hierarchical.
Then, the VEVs of Higgs fields will be hierarchical,
$\langle h^{u,d}_3 \rangle \ll \langle h^{u,d}_2 \rangle \ll \langle h^{u,d}_1 \rangle$,
and the quark masses are hierarchical.
Or, one can interpret that the effective Yukawa couplings for quarks
are (unified) gauge couplings multiplied with the hierarchical Higgs mixings.
In this choice, one of mass eigenvalues of lepton is hierarchically small,
but two eigenvalues are degenerate due to the symmetricity of the matrix.

To obtain the VEVs of singlets $s_i$, one needs brane-localized terms.
The hierarchy of the VEVs may be related to the volume of each torus,
since the brane-localized terms needs to be generated by Wilson line operators.

In order to obtain a phenomenological model, we have to break up and down
symmetry for quark masses, and flavor mixings have to be introduced.
It can be realized when brane-localized terms are introduced
as usual construction of a trinification model.
Also we need to break the degeneracy of charged-lepton masses.
It can be done by introducing vector-like brane-localized fields
which is mixed with the bulk fields,
and the second generation of charged-lepton is replaced to the brane-localized field.

We note that the bulk interaction also contains
colored Higgs couplings, e.g. $q q h^C$, $q\ell \bar h^C$ etc.
They can generate the dimension five operator for proton decay.
Actually, colored Higgs masses are also generated when
the singlet components $s_i$ acquire VEVs.
When the VEVs are hierarchical, there are light colored Higgs fields,
and it causes a rapid proton decay.
Even if we add mass terms of colored Higgs fields,
the colored Higgs couplings are still dangerous
when the fermion mass hierarchy comes from doublet Higgs mixings
keeping the original coupling is unified to the gauge coupling as noted above.
Since the colored Higgs fields do not have hierarchical mixing,
the colored Higgs couplings does not have flavor suppression
and it generate too large nucleon decay amplitudes.
To avoid it, the bulk fields should be replaced with a brane field
by introducing vector-like brane matter.
Or, the colored Higgs fields can be projected out by adopting
additional orbifold projection.
In the latter case, however, some of the generations are also projected out.
For example, suppose that all $h_i^C$ are projected out.
When the discrete charge of $h_3^C$ is not zero,
the discrete charge of one of $q_1$ and $q_2$ component can not be zero
due to the conservation of the discrete charge in the coupling $q_i q_j h_k^C$.
%two of the quark doublets must be also projected out.
As a result, two of the quark doublets $q_i$ are projected out.
Then, the quark mass matrix from bulk interaction has to be (at most) rank 1.
Similarly, due to the coupling $u^c e^c h^C$,
at least one of the right-handed charged-lepton is projected out.
Therefore, if the colored Higgs fields are projected out by orbifold
in this branch, we obtain as a consequence  the first generation of
quarks and leptons to be always massless when the brane-localized
terms are suppressed.

%The brane localized term, however, can be suppressed by a choice of extra $U(1)$ symmetries
%remained in 4D.
%The bulk interaction $Q_1^c Q_2^c Q_3^c$ includes
%a $R$-parity violating term $u^c_i d^c_j d^c_k$, which is much dangerous
%to a rapid proton decay.
%Therefore, the model is not phenomenologically viable unless the right-handed quarks
%are mixed with brane-localized fields and a light eigenstate for the quark is almost the brane fields.
%Namely, the zero modes of right-handed quarks should not be our fermions.

We also note that $E_7$ adjoint has one flavor of trinification matter.
Therefore, we obtain the three family trinification model in $E_7$ bulk symmetry
using the case 3 solution.
In that case, the Yukawa matrix is anti-symmetric, and thus the two eigenvalues are
degenerate, and one eigenvalue is zero.

\subsection{SM decomposition}

As we have noted, under the Pati-Salam branch,
$SU(4)_c$ sextet $({\bf 6},{\bf 1},{\bf 1})$ can include the right-handed down-type quark.
Similarly, when the PS symmetry is broken down to SM gauge symmetry,
 $({\bf 6},{\bf 2},{\bf 2})$, $({\bf 15},{\bf 1},{\bf 1})$, $({\bf 1},{\bf 2},{\bf 2})$,
and $({\bf 1},{\bf 1},{\bf 3})$ representations contain the quark
doublet, right-handed up-type quark, lepton doublet, and
right-handed charged lepton, respectively.
Therefore, one more family can be included in the adjoint representation.
In $E_7$ and $E_8$, the number of family is three and five, respectively,
which can be found if we see the branches,
$E_7 \to SU(8) \,({\rm or}\ SU(6) \times SU(3)) \to SU(5) \times SU(3) \times U(1)$ and
$E_8 \to SU(5) \times SU(5)$:
\begin{eqnarray}
{\bf 133} &=& ({\bf 24},{\bf 1})+ ({\bf 10},\overline{\bf 3})+(\overline{\bf 10},{\bf 3})
+ ({\bf 5},\overline{\bf 3})+(\overline{\bf 5},{\bf 3}) +
({\bf 5},{\bf 1})+(\overline{\bf 5},{\bf 1}) + ({\bf 1},{\bf 8})+ ({\bf 1},{\bf 1}), \\
{\bf 248} &=& ({\bf 24},{\bf 1})+ ({\bf 10},\overline{\bf 5})+(\overline{\bf 10},{\bf 5})
+ ({\bf 5},{\bf 10})+(\overline{\bf 5},\overline{\bf 10})  + ({\bf 1},{\bf 24}).
\end{eqnarray}
The discrete charge assignments for the SM decomposed representations
in the adjoint of $E_7$ and $E_8$ are given in Appendix.

One can find that if we assign all of three quark doublets to have zero modes
for the case 1 solution ($x_1+x_2+x_3 =0$, for example),
$({\bf 1},{\bf 2})_{\pm 1/2}$ components remain massless in the vector multiplet
for both $E_7$ and $E_8$ cases.
Therefore, in the case 1 solution, $SU(3)_L (\supset SU(2)_L)$ symmetry always remains in 4D.
As a consequence, when the gauge symmetry is broken down to SM and the
global or gauged flavor symmetry is completely broken to $U(1)$'s,
(at least) one of generation of quark
 is massless in the limit where the brane-localized couplings are zero.
%for a Yukawa interactions originated from bulk interaction
%when $E_7$ and $E_8$ is broken down to SM by orbifold.
%
In the case where the bulk gauge symmetry is $E_8$
and the trinification $G_{333}$ or
$E_6$ symmetry remains in 4D, all three generation can become massive as we have seen.
(In this case, at least $SU(3)_L$ symmetry remains in 4D.)
%
%In these two cases, the first generation can become massive.
This is because $x_1 + x_2 + x_3 = 0$ is satisfied (up to
permutation of $SU(5)$ flavor index) for these two branches.
However, as we have noted, the dangerous dimension five operators
can be generated.
Therefore,
one of generation of quark fields
should be replaced to a brane-localized field.
Or, the colored Higgs fields should be projected out
by additional orbifold condition.
In any cases, the mass of first generation should be small
when the flavor symmetries are completely broken and the
brane-localized couplings are suppressed.

%and (at least) one of generation of quark fields
%should be replaced to a brane-localized field.
%Or, the colored Higgs fields should be projected out
%by additional orbifold condition.
%Then, some of the quark fields are also projected out.
%
%since all three generations of right-handed
%quarks have zero modes in each $\Sigma_i$ in that case,
%$R$ parity violating terms are generated from the bulk gauge interaction.
%
%Therefore, it is better to projected out one of the generation by
%orbifold.
%If it is the situation, one of the generation is always massless
%when the brane-localized terms are suppressed.

Here, we consider the case where the 4D symmetry is the standard model,
and three generations of right-handed quarks
$u^c$ are obtained from bulk.

\bigskip

{\bf Example 9}:
Let us obtain the case 1 solution considered in section 4
in $T^2/\mathbb{Z}_6 \times T^2/\mathbb{Z}_3$ orbifold in $E_7$ bulk symmetry.
The discrete charges are assigned as
\begin{eqnarray}
&&(x_1,x_2,x_3)_{{Z}_6} = (0,1,2), \quad (x_1,x_2,x_3)_{{Z}_3} = (1,1,1), \quad
a_{{Z}_6} = 1, \quad a_{{Z}_3} = 0, \\
&&(k_1,l_1,m_1) = (1,2,3), \quad (k_2,l_2,m_2)= (1,1,1).
\end{eqnarray}
Then, the 4D symmetry is $G_{\rm SM} \times U(1)^3$, and the zero modes are
\begin{equation}
\begin{array}{c|c}
\Sigma_1 &  q_2, u_1^c, e^c_3 \\
\Sigma_2 &  q_3, u_2^c, \bar \ell^{23} \\
\Sigma_3 &  u_3^c, \bar \ell^{13}, \bar d^{c23}
\end{array}
\end{equation}
The three right-handed up-type quarks are the zero modes of bulk fields.
The bulk interaction includes
\begin{equation}
q_2 u^c_3 \bar \ell^{23} + q_3 u_1^c \bar \ell^{13} + u^c_2 e_3^c \bar d^{c23}.
\end{equation}

The fields $\bar \ell^{23}$ and $\bar \ell^{13}$
can be considered as up-type Higgs fields.
When the Higgs fields acquire VEVs, two eigenvalues of up-type quarks
become non-zero.
%As we have described in the case of trinification,
The eigenvalues can be hierarchical when the Higgs mixing
is small due to the suppression of couplings in the brane-localized interaction.
In this example, one can construct the model in 8D, and thus,
the scalar component $\Sigma_3$ can be chosen not to be gauge fields,
but, $\Sigma_1$ and $\Sigma_2$ can be considered the gauge fields with extra dimensional
components.
Then, the light linear combination of the Higgs fields
can be almost $\bar \ell^{23}$ rather than $\bar \ell^{13}$,
since the coupling with $\bar \ell^{23}$
will be
generated by a Wilson line operator
and it is expected to be small
\cite{Gogoladze:2007nb}.

In the bulk $E_8$ case, we can find a similar solution as Example above.
However, it is rare to obtain both up- and down-type quarks as well
as charged leptons to get masses if the symmetry is broken down to
SM, because many of the components are projected out in that case.
If some of the symmetry remains like $SU(4)_c \times SU(2)_L \times
U(1)_R$ symmetry or $SU(3)_c \times SU(2)_L \times SU(2)_R \times
U(1)_{B-L}$ symmetry, one can obtain their Yukawa couplings (namely,
top, bottom, and tau Yukawa couplings) from the bulk gauge
interaction. As we have noted, in such 4D symmetries, all three
generations of left-handed quarks cannot have zero modes when bulk
flavor symmetry is completely broken.

As we have seen in section 4, in the cases 2 and 4 where $SU(2)$ flavor symmetry remains in 4D,
two of the mass eigenvalues are degenerate in the Pati-Salam branch
(namely, the mass matrix is rank 1 or eigenvalues are degenerate even if the rank is 2).
This is because of the restriction of discrete charge assignment $2e=0$ in the Pati-Salam branch.
When the 4D symmetry is $G_{\rm SM}$, we have additional examples
where the mass matrix is rank 2 and the eigenvalues are not degenerate.
Here we give an example in the case where gauged $SU(2)$ flavor symmetry remains in 4D.

\medskip

{\bf Example 10}:
Let us obtain the case 4 solution considered in section 4
in $T^2/\mathbb{Z}_6 \times T^2/\mathbb{Z}_3$ orbifold in $E_8$ bulk symmetry.
The discrete charges are assigned as
\begin{eqnarray}
&&(x_1,x_2,x_3)_{{Z}_6} = (1,1,2,3,4), \quad (x_1,x_2,x_3)_{{Z}_3} = (1,1,1,0,0), \quad
a_{{Z}_6} = 1, \quad a_{{Z}_3} = 0, \\
&&(k_1,l_1,m_1) = (1,2,3), \quad (k_2,l_2,m_2)= (1,1,1).
\end{eqnarray}
Then, the 4D symmetry is $G_{SM} \times SU(2) \times U(1)^2$, and the zero modes are
\begin{equation}
\begin{array}{c|c}
\Sigma_1 &  q_1, q_2, e^c_3, \ell_{35} \\
\Sigma_2 &  q_3, u_1^c, u_2^c, \bar \ell^{13}, \bar \ell^{23} \\
\Sigma_3 &  u_3^c, \bar \ell^{12}, \bar d^{c13}, \bar d^{c23}, S^5_1, S^5_2
\end{array}
\end{equation}
%
%The three right-handed up-type quarks are the zero modes of bulk fields.
The bulk interaction includes
\begin{equation}
(q_1 u^c_2 + q_2 u^c_1) \bar \ell^{12} + (q_1 \bar \ell^{13} + q_2 \bar \ell^{23}) u^c_3
+ (u_1 \bar d^{c13} + u_2 \bar d^{c23}) e_3^c
+ \ell_{35} (\bar \ell^{13} S^5_1 + \bar \ell^{23} S^5_2).
%
%q_2 u^c_3 \bar \ell^{23} + q_3 u_1^c \bar \ell^{13} + u^c_2 e_3^c \bar d^{c23}.
\end{equation}
The fields $\bar d^{c13},\bar d^{c23}$ are colored Higgs, and $\bar \ell^{12,13,23}$
are up-type Higgs fields.
The field $\ell_{35}$ can be considered as a lepton doublet,
and then, $S^5_1$ and $S^5_2$ corresponds to right-handed neutrinos.
Note that $q\ell \bar d^c$ coupling is forbidden.
The field $\ell_{35}$ can be also considered as a Higgs doublet,
and $S^5_1$ and $S^5_2$ form a $SU(2)$ flavor doublet.
When $S^5_1$ acquires a VEV, the flavor $SU(2)$ symmetry is broken,
and $\ell_{35}$ and $\bar \ell^{13}$ become massive.
Then, the quark mass matrix from the bulk interaction is
\begin{equation}
(q_3, q_1, q_2) \left( \begin{array}{ccc}
                0 & 0 & 0 \\
                0 & \langle \bar \ell^{12} \rangle& 0 \\
                \langle\bar \ell^{12} \rangle& 0 & \langle\bar \ell^{23}\rangle
            \end{array}
        \right)
\left( \begin{array}{c}
     u_1 \\ u_2 \\ u_3
    \end{array}
\right).
\end{equation}
As is the previous example,
the Higgs mixing can become small, and the two VEVs are hierarchical.
Then, two eigenvalues are hierarchical, and one eigenvalue is zero.

One can also consider a similar solution where global $SU(2)$ flavor symmetry remains.
Also, in the case of $E_8$ bulk symmetry,
the discrete charges can be assigned so that both global and gauged flavor symmetries
remain in 4D.

\section{Conclusion }

We studied the higher dimensional models
in which
gauge, Higgs and three families of matter fields
are unified in a SUSY gauge multiplet.
The gauge multiplet contains three chiral superfields $\Sigma_i$ as well as
a vector multiplet in 4D point of view.
When both Higgs and matter fields are extracted as zero modes
by orbifold projection,
the Yukawa interaction in 4D is generated from the bulk gauge interaction.

We classified the cases in which three generations
are included in the zero modes of the bulk superfields.
The bulk gauge symmetry is broken down to the 4D gauge symmetry
which contains the SM gauge group by orbifolding,
and some of non-Abelian gauged flavor symmetry can also remain in 4D.
For example, $E_8$ has a subgroup $SO(10) \times SU(4)$,
and thus in the $SO(10)$ basis, there can be maximally
$SU(4)$ gauged flavor symmetry.
Furthermore, since there are three chiral superfields,
there is a global symmetry which originates from $R$ symmetry in the bulk.
If the gravity is taken into account, the $R$ symmetry is also gauged,
but we consider it as a global symmetry in a flat limit.
Therefore, to obtain three generations,
there are cases where the gauged or global flavor symmetry remains.
One can assign to make $SU(2)$ and $SU(3)$ flavor symmetries
remain for both global and gauged
symmetries.
Also, it can be considered that the both global and gauged flavor symmetry
is completely broken by orbifolding.
Totally, there are 5 cases to obtain three generations from the bulk fields.
We investigated each case and gave examples for the discrete charge
assignment of the orbifold.

Due to the bulk flavor symmetry,
the Yukawa structure originated from the bulk gauge interaction
is restricted.
Many of the cases, two eigenvalues are degenerate
in the limit where the couplings in brane-localized terms
are zero.
There are two situations in those cases:
Two eigenvalues are degenerate and one eigenvalue is zero.
Or, two eigenvalues are zero, and one eigenstate is massive.
Since the brane-localized terms are suppressed by a factor from the
volume of the extra dimension and the bulk interaction gives a dominant
contribution to the Yukawa couplings,
the former situation is not a good situation phenomenologically.
To make a phenomenologically viable model, one needs to mix one of the eigenstate
with a brane field,
and the brane field is a light eigenstate of our quarks and leptons.
In the latter situation, on the other hand, one can explain why only
third generation is heavy. The two other generations can become
massive by brane terms. The hierarchy between first and second
generation can be explained by the  remaining flavor symmetry.

When the non-Abelian flavor symmetry for both global and gauged symmetry
is broken by orbifold,
the degeneracy of eigenvalues can be resolved.
In that case, the discrete charge assignment is restricted,
and higher gauge symmetry often remains in 4D.
Actually, when all three quark doublets have zero modes in that case,
at least $SU(3)_L$ gauge symmetry (which contains $SU(2)_L$) remains in 4D.
In other words, if the gauge symmetry is broken down to $SU(2)_L$,
all three quark doublets can not have zero modes,
and at least one of the eigenvalues is zero.
It can interestingly explain why the first generation mass is tiny.
We gave an example where all three generations of quarks have non-degenerate
masses from the bulk Yukawa interaction in the model
in which trinification symmetry remains in 4D.
We also gave an example where two of the up-type quarks have non-degenerate
eigenvalues from the bulk Yukawa interaction with 4D standard model gauge group.

In conclusion,
the Yukawa coupling is generated from the bulk gauge interaction
when left- and right-handed matter fields as well as Higgs fields
are zero modes of the bulk fields.
We considered the cases where three generations of matter
are contained in the zero modes.
In that cases, due to the bulk flavor symmetries, the structure of
bulk Yukawa coupling is restricted
and the hierarchy of fermion masses can be explained
by a nature of extra dimensions.

\appendix
\section*{Appendix : Charge assignments for SM decomposed representations}

We obtain the discrete charge assignments for the SM decomposed representations
in the $E_7$ adjoint.
The branch $E_7 \to SU(5) \times SU(3) \times U(1)$ is useful
to arrange the representations.

For the adjoint representations for $G_{\rm SM}$ and $SU(3)$, the charges are
\begin{equation}
\begin{array}{|c|c|c||c|}
\hline
({\bf 8},{\bf 1})_0+({\bf 1},{\bf 3})_0+({\bf 1},{\bf 1})_0
 & ({\bf 3},{\bf 2})_{-5/6} & (\bar{\bf 3},{\bf 2})_{5/6} & S_i{}^j \\
\hline \hline
0 & a & -a  & x_i - x_j \\
\hline
\end{array}
\end{equation}
where $S_i{}^j$ is a $SU(3)$ adjoint representation $(i=1,2,3)$.

The matter and anti-matter representations are given as follows:

Discrete charges for the representations in $({\bf 10},\bar{\bf 3})$ and $(\overline{\bf 10},{\bf 3})$ are
\begin{equation}
\begin{array}{|c|c|c||c|c|c|}
\hline
 q_i & u_i^c & e_i^c & \bar q^i & \bar u^{ci} & \bar e^{ci}
  \\
\hline \hline
x_i & a+x_i & -a+x_i & -x_i & -a -x_i & a-x_i
  \\
\hline
\end{array}
\end{equation}

The representations in $(\overline{\bf 5},{\bf 3})$ and $({\bf 5},\bar{\bf 3})$  are
\begin{equation}
\begin{array}{|c|c||c|c|}
\hline
 d_{ij}^c & \ell_{ij}
& \bar d^{cij} & \bar \ell^{ij}  \\
\hline \hline
x_i + x_j & x_i+x_j+a &  -x_i - x_j & -x_i-x_j-a  \\
\hline
\end{array}
\end{equation}

The representations in $(\overline{\bf 5},{\bf 1})$ and $({\bf 5},{\bf 1})$  are
\begin{equation}
\begin{array}{|c|c||c|c|}
\hline
 D^c & \bar H
& \bar D^c & H  \\
\hline \hline
-x_1-x_2-x_3 -a  & -x_1-x_2-x_3 &  x_1+x_2+x_3+a & x_1+x_2+x_3  \\
\hline
\end{array}
\end{equation}

When $x_3 + a \equiv 0$ is satisfied for example, $SU(4)_c \times SU(2)_L \times U(1)_R$ symmetry remains in 4D.
When $x_3 - a \equiv 0$ is satisfied, $SU(3)_c \times SU(2)_L \times SU(2)_R \times U(1)_{B-L}$ symmetry remains.
When $x_3 + a \equiv 0$ and $2a \equiv 0$ are satisfied, Pati-Salam symmetry remains.
When $x_2 \equiv x_3 \equiv a$ and $3a \equiv 0$ are satisfied, trinification symmetry remains.
When $a\equiv 0$ is satisfied, $SU(5)$ symmetry remains.
When $x_3 \equiv 0$ is satisfied, flipped-$SU(5)$ remains.
When $x_3 \equiv a \equiv 0$ is satisfied, $SO(10)$ symmetry remains.
When $x_2 \equiv x_3 \equiv a \equiv 0$ is satisfied, $E_6$ symmetry remains.

Next, let us
obtain the discrete charge assignments for the SM decomposed representations
in the $E_8$ adjoint.
The branch $E_8 \to SU(5) \times SU(5)$ is useful
to arrange the representations.

For the adjoint representations for $G_{\rm SM}$ and $SU(5)$, the charges are
\begin{equation}
\begin{array}{|c|c|c||c|}
\hline
({\bf 8},{\bf 1})_0+({\bf 1},{\bf 3})_0+({\bf 1},{\bf 1})_0
 & ({\bf 3},{\bf 2})_{-5/6} & (\bar{\bf 3},{\bf 2})_{5/6} & S_i{}^j \\
\hline \hline
0 & a & -a  & x_i - x_j \\
\hline
\end{array}
\end{equation}
where $S_i{}^j$ is a $SU(5)$ adjoint representation $(i=1,2,3,4,5)$.

The matter and anti-matter representations are given as follows:

Discrete charges for the representations in $({\bf 10},\bar{\bf 5})$ and $(\overline{\bf 10},{\bf 5})$ are
\begin{equation}
\begin{array}{|c|c|c||c|c|c|}
\hline
 q_i & u_i^c & e_i^c & \bar q^i & \bar u^{ci} & \bar e^{ci}
  \\
\hline \hline
x_i & a+x_i & -a+x_i & -x_i & -a -x_i & a-x_i
  \\
\hline
\end{array}
\end{equation}

The representations in $(\overline{\bf 5},\overline{\bf 10})$ and $({\bf 5},{\bf 10})$  are
\begin{equation}
\begin{array}{|c|c||c|c|}
\hline
 d_{ij}^c & \ell_{ij}
& \bar d^{cij} & \bar \ell^{ij}  \\
\hline \hline
x_i + x_j & x_i+x_j+a &  -x_i - x_j & -x_i-x_j-a  \\
\hline
\end{array}
\end{equation}

From the algebra, the condition $\sum x_i + a \equiv 0$ has to be satisfied.

Similarly to the $E_7$ case,
higher symmetries remain in 4D when above conditions are satisfied.

\section*{Acknowledgments}

We thank Z. Tavartkiladze for useful discussions. This work of Y. M.
is supported in part by the grant from the US Department of Energy,
grant number DE-FG02-95ER40917, and the work of S. N. was supported
in part by the grants from the US Department of Energy,  grant
numbers DE-FG02-04ER41306 and DE-FG02-ER46140.

\end{document}